# Direct determination of momentum-resolved electron transfer in photoexcited MoS$_2$/WS$_2$ van der Waals heterobilayer

Fang Liu, Qiuyang Li, X.-Y. Zhu[*]

Department of Chemistry, Columbia University, New York, NY 10027, USA

**ABSTRACT**

**Photo-induced charge separation in transition metal dichalcogenide heterobilayers is being explored for moiré excitons, spin-valley polarization, and quantum phases of excitons/electrons. While different momentum spaces may be critically involved in charge separation dynamics, little is known directly from experiments. Here we determine momentum-resolved electron dynamics in the WS$_2$/MoS$_2$ heterobilayer using time and angle resolved photoemission spectroscopy (TR-ARPES). Upon photoexcitation in the K valleys, we detect electrons in M/2, M, and Q valleys/points on time scales as short as ~70 fs, followed by dynamic equilibration in K and Q valleys in ~400 fs. The interlayer charge transfer is accompanied by momentum-specific band renormalization. These findings reveal the essential role of phonon scattering, the coexistence of direct and indirect interlayer excitons, and constraints on spin-valley polarization.**

Monolayer transition metal dichalchogenides (TMDCs) are excellent models for the exploration of semiconductor physics at the two-dimensional (2D) limit, with potential applications in electronics, optoelectronics, and quantum devices [1]. Heterobilayers formed from stacking two TMDC monolayers may bring about new physical properties, such as moire excitons [2,3], long-lived spin valley polarization [4,5] and exciton condensation [6]. For TMDC heterobilayers, photoexcitation of one monolayer leads to electron or hole transfer to the other layer, resulting in interlayer excitons with electron and hole spatially separated across the interface [7–13]. Past experiments indicated that interlayer charge transfer occurs on ultrafast (<100 fs) time scales, regardless of relative layer orientations [11,12,14,15]. This observation is surprising since the conduction band minimum (CBM) and valence band maximum (VBM) are both located at the K/K′ points at the hexagonal Brillouin zone corners and relative orientation between the two monolayers should result in momentum mismatch. While transient absorption/reflectance spectroscopies unambiguously established ultrafast charge transfer in photo-excited TMDC heterobilayers and photoluminescence spectroscopies have confirmed the formation of interlayer excitons [7–13], there is little understanding on how momentum is conserved in interlayer charge transfer and how optically dark interlayer excitons are involved.

---

[*] To whom correspondence should be addressed. E-mail: xyzhu@columbia.edu

To understand electron transfer in TMDC heterobilayers, we examine the band structure for an angle-aligned $WS_2/MoS_2$ heterobilayer from first principle calculation by Okada et al. [16], Figure 1(a). The wavefunctions at the K/K′ point are strongly confined within each monolayer, while the Q and the M/2 valleys have mixed characters [16]. A likely mechanism may involve phonon-assisted scattering of CB electrons from the K valleys of photo-excited $WS_2$ to the energetically resonant momentum spaces (Q and M/2) with strongly mixed $WS_2/MoS_2$ characters. Here we apply TR-ARPES to probe electron dynamics in $WS_2/MoS_2$ heterobilayers with time, energy, and momentum resolutions; the latter is not possible in all optical experiments. While ARPES has been used to probe the static valence band structure of a TMDC heterobilayer, TR-ARPES is applied here for the first time to determine the dynamics following across gap photo-excitation. In contrast to our TR-ARPES work on monolayer $MoS_2$ where CB electron dynamics is confined to the K valley [17], we observe ultrafast redistribution of CB electrons to multiple momentum spaces following photo-excitation of the $MoS_2/WS_2$ heterobilayer. Specifically, our finding (i) solves a major puzzle in the field – the insensitivity of ultrafast charge separation to interlayer twist angle in TMDC heterobilayers; (ii) reveals the dominant role of phonon scattering is mediating interlayer electron transfer; and (iii) identifies the presence of a dynamic equilibrium between optically bright K-K (K′-K′) and dark K(K′)-Q interlayer excitons in angle-aligned heterobilayers.

We choose the model system of $WS_2/MoS_2$ hetero-bilayer on $SiO_2$ (285 nm thickness on n-doped Si). The use of a dielectric surface minimizes disturbance to the electronic properties of TMDCs, as we demonstrated for $MoS_2$ monolayer on the same substrate [17], Improving upon previous demonstrations of metal-assisted exfoliation [18–20], we have developed a gold-tape exfoliation technique [21] to produce single crystal TMDC monolayers and heterostructures with lateral size > 3 mm, Figure 1(b). To construct the heterobilayer, we first exfoliate a $MoS_2$ monolayer on the substrate and place a second $WS_2$ monolayer on top with a small interlayer twist angle of 5°, as determined by second harmonic generation (Fig. S1). We deposit gold electrodes on the side of the heterobilayer for grounding. The visible pump ($hv_1$ = 2.2 eV, 40 fs pulse width, s-polarized, 0.5-2 mJ/cm$^2$, spot diameter 0.5-1 mm) excites electrons from VB to CB, and the extreme UV probe (EUV, $hv_2$ = 22 eV, pulse duration <100 fs, 10 kHz, p-polarized, ~10$^9$ photons/s, spot diameter 200 μm) ionizes the electrons from both valence and conduction bands at a controlled time delay for detection by a hemispherical analyzer, Figure 1b. The ARPES with EUV maps out the valence band structure for the $WS_2/MoS_2$ heterobilayer along Γ-K and Γ-M directions, Fig. 1c and 1d. Comparisons of energy distribution curves (EDCs) from the heterobilayer to those from the monolayers show that the results are consistent with hybridization at the Γ point, not the K point (Fig. S2 - S4), in agreement with the high resolution ARPES work of Wilson et al. on $MoSe_2/WSe_2$ heterobilayers [22].

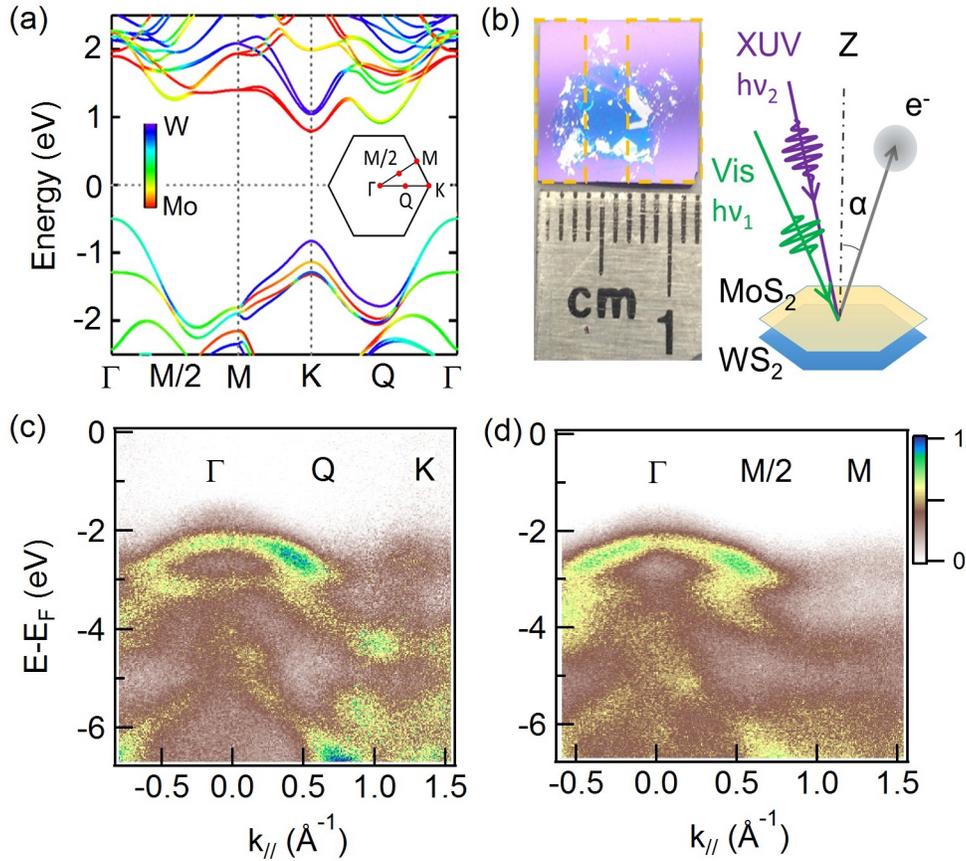

**Figure 1**. (a) Calculated band structure of $WS_2/MoS_2$ heterostructures with A-A stacking (0 degree alignment (Reprinted with permission from Ref. [16]. Copyright 2018 American Chemical Society). The color scheme indicates the wavefunction contributions from each layer. The inset is a schematic of the different points in the Brillouin zone. (b) Left: Image of the $WS_2/MoS_2$ heterostructure on 285 nm $SiO_2/Si$, with blue color from optical contrast. Gold electrodes are deposited on the side (dashed yellow line) for grounding. A zoomed image of the sample is shown in Fig. S2. Right: Schematics of the TR-ARPES experiment. A femtosecond visible pulse serves as pump and the femtosecond EUV pulse probe. The photoelectrons are detected by hemispherical analyzer at various polar angle α from surface normal for momentum resolution. Shown in the bottom panels are experimental ARPES of $WS_2/MoS_2$ heterostructure measured along (c) Γ-$K$ direction and (d) Γ-$M$ directions. All TR-ARPES and ARPES measurements are performed at room temperature.

We carry out TR-ARPES measurements around five representative positions in the momentum space, including the Brillouin zone center Γ, corner K (K′), edge M and the halfway positions of Q and M/2. The ARPES of conduction band without (Δt < 0) and with (Δt = 0) visible pump are shown in Fig. 2, with the EDCs in the bottom panels. The monolayers are intrinsically n-doped [17,23–25] and, as a result, the CBM is close to the Fermi level ($E_F$). The intrinsic n-doping is reflected in the electron population at $\Delta t < 0$ predominantly near the K point, which is predicted to be the global CB minimum [18]. When the pump pulse overlaps with the probe pulse at $\Delta t = 0$, the increases in conduction band electron population appear at K, Q, M and M/2 positions, but not at Γ. The latter is expected from its predicted high energetic location [18].

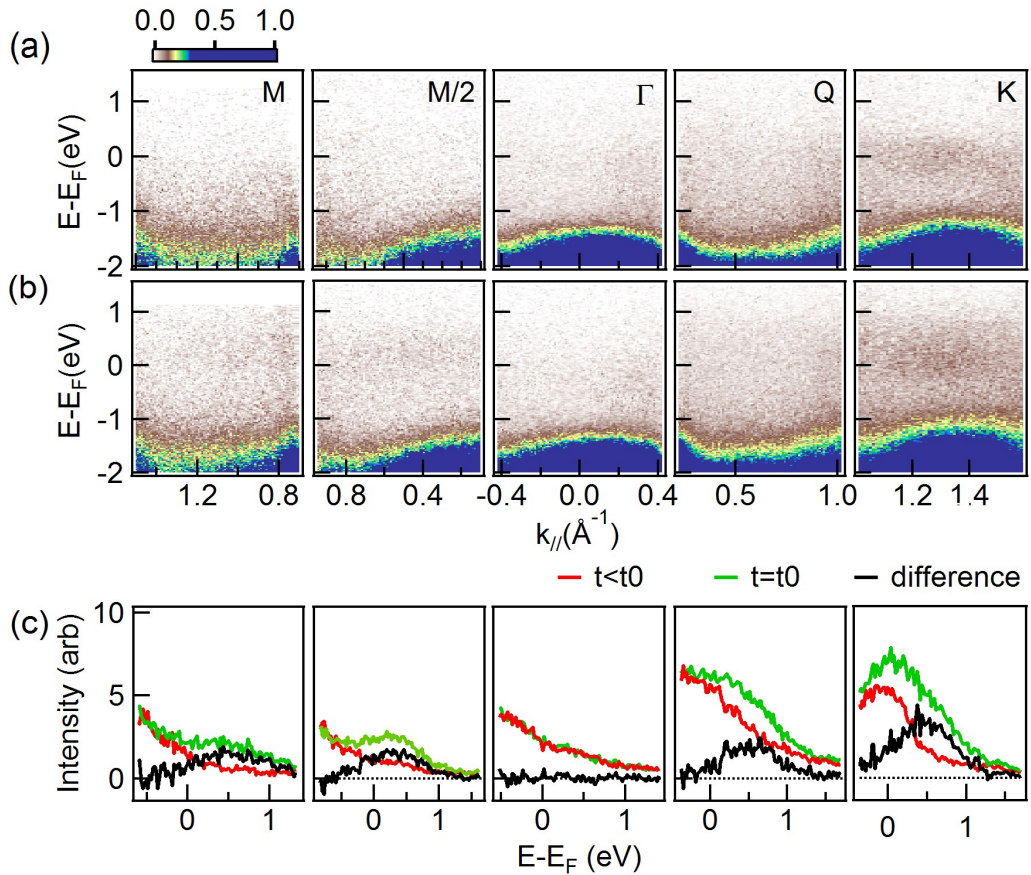

**Figure 2**. TR-ARPES of the $WS_2/MoS_2$ heterostructure (a) without (Δt<0) and (b) with the visible pump excitation (Δt=0), respectively. The ARPES spectra are collected at *M*, *M/2*, *Γ*, *Q*, *K* positions in the momentum space. The color scheme is scaled to show weak signal in the CB. (c) The corresponding electron energy distribution curve (EDC) near the CB edge, collected without (Δt<0, red) and with (Δt=0, green) the visible pump excitation, and the difference with and without pump (black). The EDCs are integrated within a ±0.15 Å$^{-1}$ window of the center momentum at each point from raw photoionization signals. The visible pump is fixed at a photon energy of 2.2 eV and the probe at 22 eV. The sample is at room temperature.

The CB electron dynamics depend on momentum, as shown in Fig. 3 for energy-integrated CB electron intensities at the K, Q, M, and M/2 positions as a function of $\Delta t$. The corresponding TR-ARPES in 2D pseudo-color plots are shown in Fig. S5. The initial pump pulse induces direct transition in the K valleys in both $WS_2$ and $MoS_2$ monolayers, as known from the much larger oscillator strength of intralayer excitons than that of the interlayer [26]. Although the limited energy resolution does not permit us to distinguish between electrons in the K valleys of $WS_2$ and $MoS_2$, respectively, early time dynamics does show a decrease in average electron energy in the K valley by ~ 100 meV on the ~0.5-1 ps time scale, as shown in Fig. S6a-b. This energy relaxation may reflect interlayer charge transfer from $WS_2$ to $MoS_2$ (via the M/2 and M intermediate states as detailed below) as well as hot electron relaxation. In contrast, there is no detectable energy relaxation on this time scale in the Q valleys in the heterobilayers, Fig. S6c-d, or in the K valleys of monolayer $MoS_2$, Fig. S6e-f.

We focus on the rise in CB populations at K, Q, M, and M/2 positions near $\Delta t=0$, as is shown in Fig. 3. The M and M/2 points undergoes a sharp rise followed by fast decay to zero. We fit the M and M/2 dynamics with an instantaneous rise with a single exponential decay, convoluted with Gaussian function (black curves in Figure3, detailed in supplementary information). The variance of the Gaussian, $\sigma = 140 \pm 30$ fs, represents an estimated upper limit in experimental time constant of ~70 fs ($1/2\ \sigma$), and indicates that interlayer electron transfer from the K valley of $WS_2$ to the M/2 and M momentum positions occur with $\tau_M \leq 70$ fs. The M/2 position has strong interlayer electronic coupling, which is also energetically close to the K valleys of $WS_2$. The population at the M point can by mediated by the M/2 point and momentum conservation in each process requires the participation of a phonon, likely a longitudinal optical (LO) phonon, via the efficient Fröhlich scattering mechanism. In most polar semiconductors, Fröhlich scattering occurs on the ultrafast time scale of a few to a few tens femtoseconds [27]. A TR-ARPES study on bulk $MoS_2$ reveals inter-valley scattering from the initially populated *K* valley to the *Σ* valley in $\leq 50$ fs [28], in excellent agreement with the present finding in the MoS$_2$/WS$_2$ heterobilayer. Following the ultrafast rise, the electron populations decay with single exponential lifetime of $\tau_{Md} = 0.7\pm0.1$ ps at both M and M/2. As we discuss below, this decay can be attributed to the transfer of electron population to the lower-lying Q (mixed WS$_2$ and MoS$_2$) and K (MoS$_2$) valleys. Auger recombination at the initial high excitation densities, as well as other nonradiative charge recombination channels, may also contribute to the 0.7 ps decay.

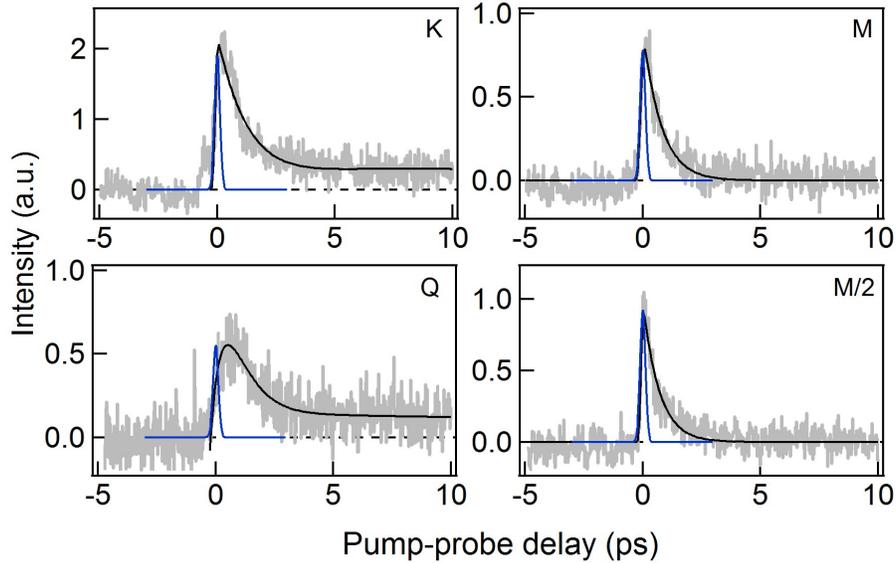

**Figure 3**. CB electron dynamics at *K*, *Q*, *M*/2, and *M* points as a function of pump probe delay. The integrated conduction band photoelectron intensities are shown as grey curves, with intensity at $\Delta t < 0$ set to zero. The solid black curves are kinetic fits and the blue curves (scaled to experimental peak intensity) are convoluted Gaussians ($\sigma = 140$ fs) representing laser pump-probe time resolution up to 70 fs ($^1/_2\,\sigma$). The dashed lines show zero intensity levels. The CB electron intensities are integrated from -0.6 eV to 2.5 eV, covering the energy window in which the CB electrons are detected. The corresponding TR-ARPES in 2D pseudo-color plots are shown in Fig. S5.

In contrast to the fast formation and decay of M/2 and M electrons, CB electron signal at K or Q in Fig. 3 has a delayed component in the rise. The kinetics of electron population at K can be fit with two components, including a prompt formation from initial excitation ($\tau_{r1} \leq 70$ fs) and the delayed formation ($\tau_{r2} = 0.6\pm0.3$ ps) which may be attributed to inter-valley back scattering from M and M/2. The delayed formation component with $\tau_{r2} = 0.6\pm0.3$ ps is more significant in Q, which may be attributed to inter-valley scattering from K and/or M and M/2. Following the rise, the decay of Q and K electrons can be described by bi-exponentials, with time constants for the fast and slow channels of $\tau_{d1} = 1.2 \pm 0.5$ ps and $\tau_{d2} \geq 40$ ps. The slow decay channel is assigned to the radiative and nonradiative recombination. The fast channel dominates initially and is likely a result of manybody scattering, such as Auger recombination, at the initial high excitation density. The majority of CB electrons reside at the K point, which is predicted to be the global conduction band minimum (CBM). The Q point lies close in energy (see Fig. 1a). Thus, the similarity in the kinetic profiles at the K and Q points at longer times reveals a dynamic equilibrium.

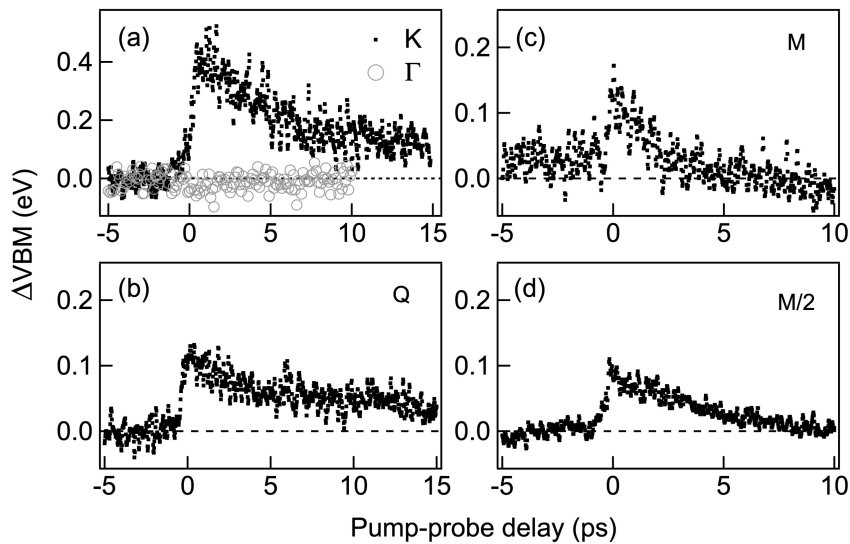

**Figure 4**. Dynamics of VBM renormalization at different momentums at a function of pump probe delay time $\Delta t$. The shift of VBM position is obtained from linear extraction at the band edge at each momentum position. The band gap renormalization ($\Delta$VBM) is the largest at $\Delta t \sim 0$ and decreases as carrier recombination proceeds with time.

Concurrent with CB dynamics, we probe the VBs that reveal band renormalization from charge carriers in each valley. The band renormalization reflects the strong many-body Coulomb interactions due to quantum confinement and reduced dielectric screening in 2D, as demonstrated in photo-excited monolayer $MoS_2$ by TR-ARPES [17,29]. While photoinduced hole populations cannot be determined due to small changes in VB signal, the shift of the VBM can be clearly resolved by linear extrapolation near the high energy cutoff (EDC in 2D pseudo color plots are shown in Fig. S7). The n-doping pins the CBM to the Fermi energy and band renormalization appears almost exclusively as upshifts in the VBM [17,29]. Band renormalization in 2D semiconductor systems is a function of momentum and depends specifically on carrier population in each valley [30]. Fig. 4 shows the time-dependent relative shifts ($\Delta$VBM) at $\Gamma$, K, Q, M/2, and M positions. Following optical excitation, band renormalization is observed at the K, Q, M/2 and M, and there are long-lived ΔVBM at K and Q points only. Such momentum-specific band renormalization is consistent with the observation of CB electrons at the same momentum spaces. Surprisingly, band renormalization is negligible at the Γ point (grey circles in Fig. 4a), despite the fact that the Γ point is the global VBM in the static $WS_2/MoS_2$ heterobilayer [31]. At high excitation densities ($10^{12-13}/cm^2$), the larger band renormalization of VB at K than that at Γ may change the energy ordering between Γ and K, leading to the accumulation of holes in the K/K' valley. This finding emphasizes the necessity of viewing the band structure of a photo-excited 2D semiconductor from a dynamic perspective.

TR-ARPES provides direct momentum-resolved measurement for the dynamic pathways in the photoexcited $WS_2/MoS_2$ heterobilayer, Fig. 5. Following the initial cross-gap excitation in the K valleys, the CB electrons scatter into Q, M and M/2 positions within ~70 fs, likely from efficient LO phonon scattering. In contrast to the layer localized K valleys, the strongly mixed CB at Q and M/2 facilitate interlayer electron transfer. On time scales of ~ 0.6 ps, the CB electrons at the M/2 and M positions are scattered to the lower energy K and Q valleys. The CB electrons in the K and Q valleys undergo recombination with VB holes on longer time scales of ≥ 40 ps. During this time, the bright K-K (K′-K′) excitons are in dynamic equilibrium with the dark K(K′)-Q excitons. The strong interlayer electronic coupling at Q and M/2 valleys and the ubiquitous involvement of LO phonon scattering may explain the ultrafast charge transfer in TMDC heterobilayers regardless of twist angles [8].

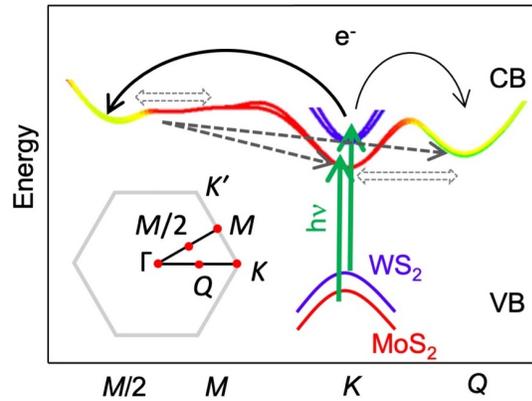

**Figure 5**. Schematics of intervalley electron scattering in the CB. The visible pump beam excites the electron from VB to CB in the *K* valleys (green arrows), followed by ultrafast scattering (≤ 70 fs) of electron from K to M, M/2 and Q valleys (black arrows). The dashed arrows represent subsequent electron scattering (~0.4 ps) from M/2 and M to K and Q. The double-ended and dashed arrows represent dynamic equilibria.

While spin-valley specific excitation at K/K′ points can be achieved in TMDC monolayers and heterostructures with circularly polarized light [32,33], our TR-ARPES experiments with reveal ultras depolarization of CB electrons in the K and K′ valleys following circularly polarized excitation. This is understood from the small spin-orbital splitting (~10s meV) of the CB at K/K′ points [34], resulting in approximately equal probability of electron scattering from M/2 and M to K and K′ valleys. The dynamic equilibrium between Q and K/K′ further scrambles the CB spin-valley polarization. A recent TR-ARPES study of bulk $WSe_2$ also reveals that CB valley polarization is lost in ~100 fs [35]. In contrast to the CB, spin-orbit energy splitting in VB is an order of magnitude larger (100s meV) [34]. Therefore, spin polarized hole scattering from K/K′ to the opposite K′/K valleys will need to overcome a large energy offset. Holes in the VB may maintain a high degree of spin-valley polarization. Although unpolarized CB electrons occupy both K/K′ valleys, only electrons at the same valley as holes can give rise to optically bright transitions. As a result, spin valley polarized holes may be solely responsible for the circularly polarized light emission/absorption/reflection, as is measured in previous optical spectra [4,5,34]. In optical spectroscopy measurements, CB electrons at the Q and the spin-opposite K′(K) valleys may act as dark reservoirs, albeit in competition with phonon assisted radiative or nonradiative recombination.

## Acknowledgement

XYZ acknowledges the National Science Foundation (NSF) grant DMR-1809680 for supporting the TR-ARPES experiments. XYZ acknowledges the Center for Precision Assembly of Superstratic and Superatomic Solids, a Materials Science and Engineering Research Center (MRSEC) through NSF grant DMR-1420634, the Columbia Nano Initiative and the Vannevar Bush Faculty Fellowship through Office of Naval Research Grant # N00014-18-1-2080 for the purchase of the laser equipment. FL acknowledges support by the Department of Energy (DOE) Office of Energy Efficiency and Renewable Energy (EERE) Postdoctoral Research Award under the EERE Solar Energy Technologies Office administered by the Oak Ridge Institute for Science and Education (ORISE). ORISE is managed by Oak Ridge Associated Universities (ORAU) under DOE contract number DE-SC00014664. All opinions expressed in this paper are the author's and do not necessarily reflect the policies and views of DOE, ORAU, or ORISE. We acknowledge

Supplementary Information

**Direct determination of momentum-resolved electron transfer in photoexcited MoS$_2$/WS$_2$ van der Waals heterobilayer**

Fang Liu, Qiuyang Li, X.-Y. Zhu[1]

Department of Chemistry, Columbia University, New York, NY 10027, USA

**Data Analysis**

Fit of the CB electron dynamics

The dynamics of CB electron at the M and M/2 positions are fit with the promt rise and a single exponential decay, convoluted with a Gaussian function to represent laser pump-probe cross correlation:

$$y = C + A\exp\left(-\frac{t-t_0}{\tau} + 0.125\left(\frac{\sigma}{\tau}\right)^2\right)\left(1 + \mathrm{erf}\left(\sqrt{2}\frac{t-t_0}{\sigma} - \frac{\sigma}{2\sqrt{2}\tau}\right)\right)$$

C is a constant offset t, A is intensity scale, $t_0$ is time zero, $\tau$ is the decay lifetime, and $\sigma$ is the variance of Gaussian function. The variance of the convoluting Gaussian, $\sigma = 140 \pm 30$ fs, corresponds to the sharpest rise observed in experiment, and represents an estimated experimental time resolution up to ~70 fs ($^1/_2$ $\sigma$). This estimation of $\sigma$ represents an limit and the actual could be better than this value.

The dynamics of the CB electron at K and Q valleys are fit with two components: one prompt rises and one delayed rise, followed by a double exponential decay, and convoluted with Gaussian function for the laser pump-probe cross correlation, with $\sigma$ (=140 fs) as obtained from fits to kinetics at the M and M/2 points detailed above. The fit function is:

$$y = C + A\left(\exp\left(-\frac{t-t_0}{\tau_{rise}}(t-t_0) + \frac{\sigma^2}{2\tau_{rise}^2}\right) - B\exp\left(-\frac{t-t_0}{\tau_1} + \frac{\sigma^2}{2\tau_1^2}\right) - (1-B)\exp\left(-\frac{t-t_0}{\tau_2} + \frac{\sigma^2}{2\tau_2^2}\right)\right) + D\left(B\exp\left(-\frac{t-t_0}{\tau_1} + 0.125\left(\frac{\sigma}{\tau_1}\right)^2\right)\left(1 + \mathrm{erf}\left(\sqrt{2}\frac{t-t_0}{\sigma} - \frac{\sigma}{2\sqrt{2}\tau_1}\right)\right) + (1-B)\exp\left(-\frac{t-t_0}{\tau_2} + 0.125\left(\frac{\sigma}{\tau_2}\right)^2\right)\left(1 + \mathrm{erf}\left(\sqrt{2}\frac{t-t_0}{\sigma} - \frac{\sigma}{2\sqrt{2}\tau_2}\right)\right)\right)$$

---

[1] To whom correspondence should be addressed. E-mail: xyzhu@columbia.edu

C is a constant offset; $t_0$ is time zero; A is intensity scale for the slow rise with time constant $\tau_{rise}$; D is the intensity scale for the instantaneous rise component; B and 1-B are the ratio of the two double exponential decay components; $\tau_1$ and $\tau_2$ are the double exponential decay lifetimes, and σ (=140 fs) is the variance of the Gaussian function.

Estimation of the energy difference between K and Q

In principle, one may estimate the energy difference between K and Q from the Fermi Dirac distribution:

$$n = \frac{1}{\exp\left(\frac{E}{k_B T}\right) + 1}$$

Using the intensity ratio of CB electrons at the K point and Q point at long delay times (averaged over 10-15 ps), assuming that at this time the electrons reside at the minimum of the K and Q points, and reach a quasi-equilibrium. This estimation gives a difference of $\Delta CBM_{Q-K}$ ~ 20 meV, which is lower than the predicted value of 130 meV without carriers [18]. The assumption of same EUV ionization cross section of CB electrons from the K and Q valleys may overestimate the electron population in the latter, although the calculated ionization cross sections at K and Σ points for monolayer MoS$_2$ are similar [35] (see Fig. S8). The detection efficiency can be different for the K and Q point electrons, thereby the measured CB electron populations can be affected by the different ionization efficiencies. The bands at K and Q are mostly composed of the metal d orbital electrons. In MoS$_2$ monolayers, the K point CBM is predominantly composed of $d_{z^2}$ while the Q point CBM is mostly $d_{x^2-y^2}$, with a small proportion of S 3p, as is shown in Figure S7.[3] At photoionization photon energy of 22 eV, the Mo 4d has a much larger cross section than S 3p orbital.[2] Moreover, the detection scheme is using a p polarized XUV beam, with incidence angle of 10° between XUV and sample surface normal. At this geometry, the $d_{z^2}$ has a larger photoionization cross section than $d_{x^2-y^2}$.[5] As a result, the CB electrons at K point is expected to have a larger detection efficiency than the Q point. Therefore, the actual energy difference of the K and Q points is expected to be even smaller than the estimated $\Delta CBM_{Q-K}$ ~ 20 meV. This is consistent with the previous calculation, showing that the band renormalization is more significant in Q than K that shifts the Q to much lower energy in monolayers.[6]

Another reason for the difference between our estimated $\Delta CBM_{Q-K}$ and calculated one may come from differential band renormalization. Upon electron doping, the $\Sigma$ point in each monolayer, akin to the Q point in a heterobilayer, is predicted to shift more significantly than the K point [36].

**Supplementary Figures**

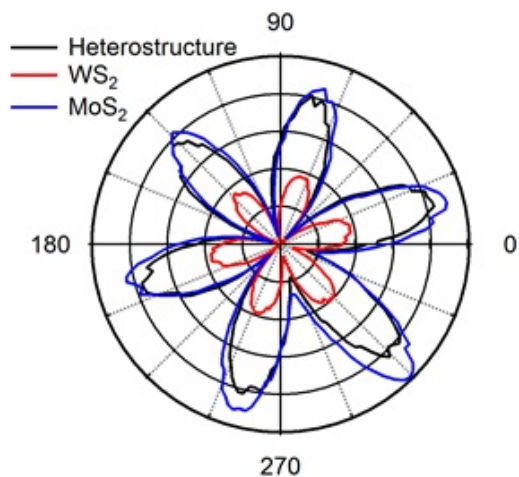

Figure S1. Determination of $WS_2/MoS_2$ crystal orientation via polarization-resolved SHG, in which the sample orientation is rotated relative to input laser polarization and collection polarization. The correlation between the phase and $\Gamma$-K (zigzag) direction is calibrated with fit from SHG data obtained from CVD grown $MoS_2$ monolayers. The twist angle between $MoS_2$ and $WS_2$ monolayers is 5°.

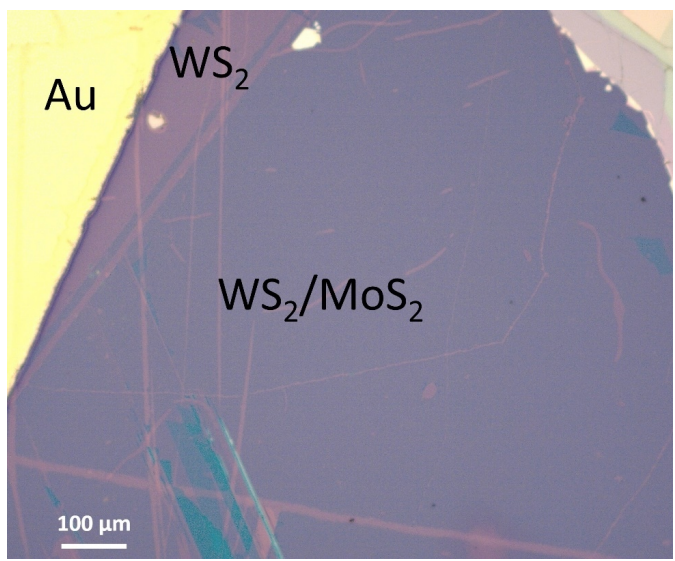

Figure S2. Zoomed in region of the $WS_2/MoS_2$ heterostructure for ARPES measurement.

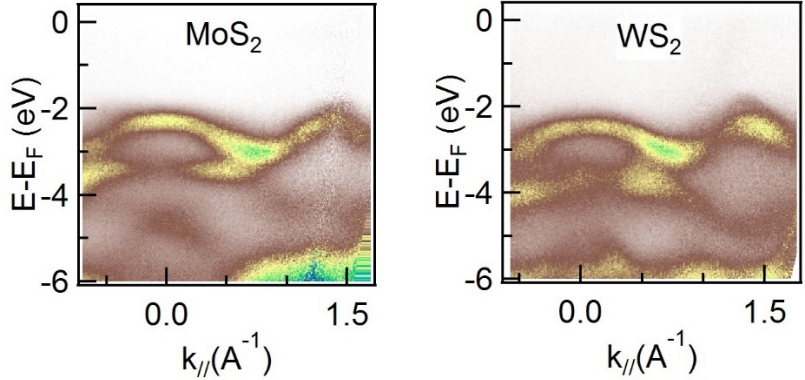

Figure S3. EUV-ARPES spectra of monolayer $MoS_2$(left), monolayer $WS_2$(right) on $SiO_2$/Si surface. Although the spin splitting bands at K point cannot be clearly resolved, the monolayer APRES is different from that of the heterostructues in the main text. The $MoS_2$ ARPES is consistent with that reported for $MoS_2$ monolayer supported on $SiO_2$ substrate[1].

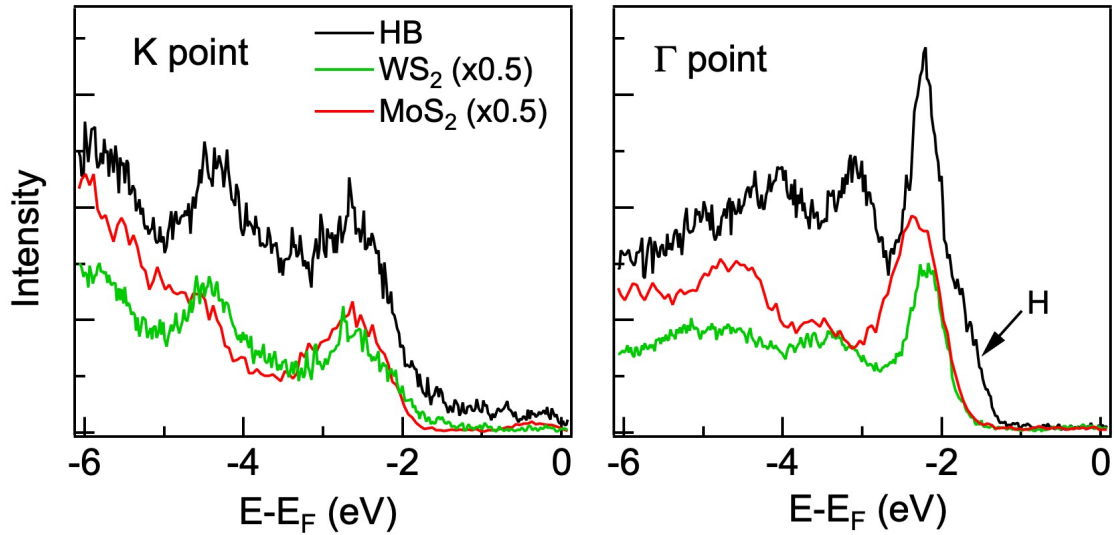

Figure S4. EDC of ARPES from $MoS_2$/$WS_2$ heterobilayer (black), $MoS_2$ monolayer (red) and $WS_2$ monolayer (green) on $SiO_2$/Si substrates. The left shows spectra at the K point and right at the Γ point. Each spectrum was integrated in a momentum window of ±0.1 Å$^{-1}$. For the K point on the left, the heterobilayer spectrum can be well described by a weighted sum of the spectra from the monolayers. This is not the case at the Γ point where the heterobilayer spectrum deviates significantly from the weighted sum of the monolayer spectra, consistent with the presence of hybridization. This is more evident at the high energy end, where an extra feature (labeled "H") is observed and can be attributed to of the VBM of the hybridized state. This is consistent with the higher resolution work of Wilson et al. on $MoSe_2$/$WSe_2$ [ref. 22].

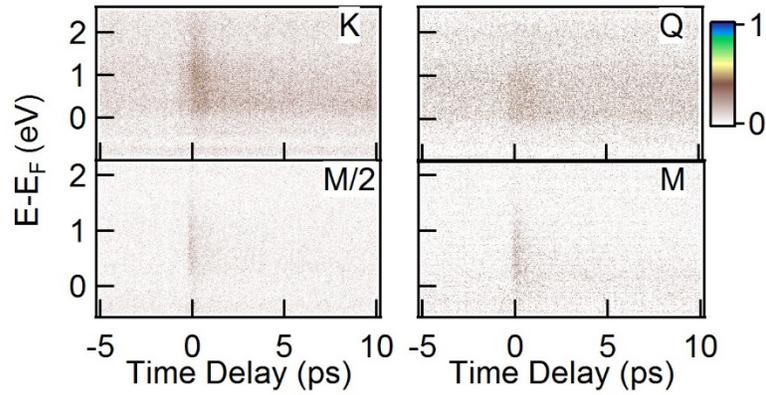

Figure S5. 2D pseudo-color (intensity) plots of EDC spectra collected at K, Q, M/2, and M points of the Brillouin zone at different momentums as a function of pump probe delay time Δt. The CB electron intensities shown in Figure 3 of main text are integrated from -0.6 eV to 2.5 eV, covering the entire energy range in which the CB electrons are detected.

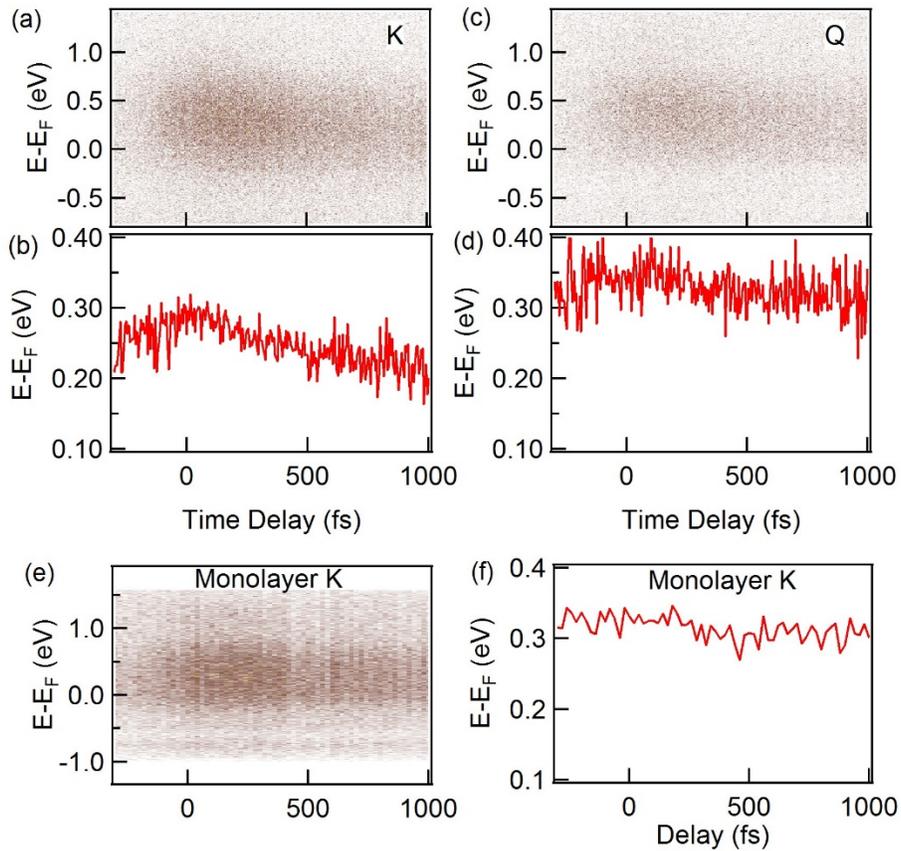

Figure S6. (a)(c)Early time TR-ARPES of K and Q points measured with the polar angle set to collect photoemission from both of them simoutaneously. (b)(d) The intensity-averaged CB electron energy extracted from panel a and c, showing the average energy is relaxed at the K point by approximately 0.1 eV, while the energy of Q point stays constant. The relaxation of CB electrons at K point of monolayer $MoS_2$ is displayed in panel (e) and (f). For comparision, there is little energy relaxation for CB electrons in the K valley.

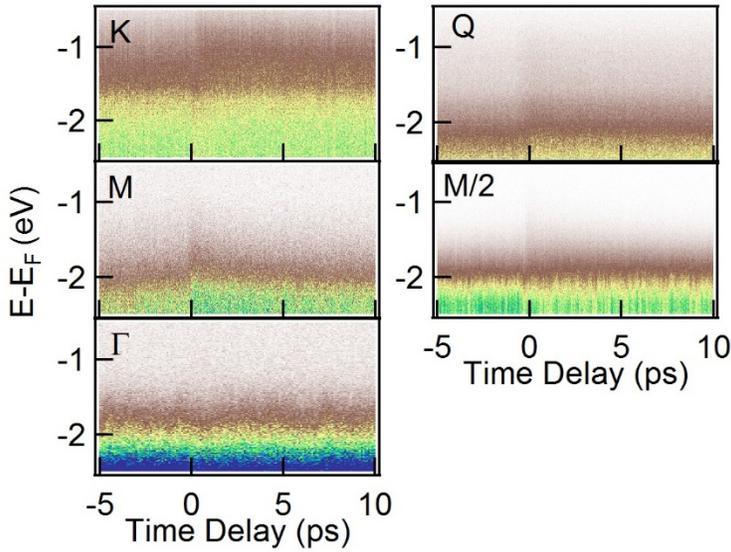

Figure S7. 2D pseudo-color (intensity) plots of EDC spectra collected at K, Q, M/2, M, and Γ points as a function of pump probe delay time Δt. The photoelectron signals are integrated within a 0.3 Å$^{-1}$ window of the center momentum.

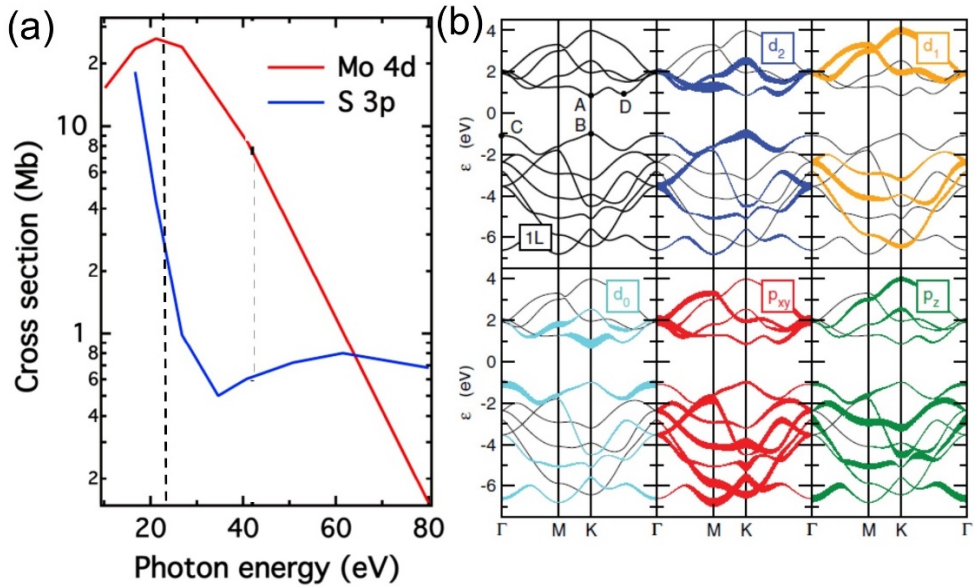

Figure S8. (a) Calculated cross sections for photoionization of Mo 4d and S 3p orbitals using XUV with different photon energy.[2] Reproduced by permission of American Physical Society. (b) Band structure and orbital character of MoS$_2$ monolayer. The thickness of the bands represents the weight of different orbitals, including: Mo 4d orbitals. $d_0 = d_{z^2}$, $d_1 = d_{xz}, d_{yz}$, $d_2 = d_{x^2-y^2}$, $d_{xy}$. and S 2p orbitals. $p_{xy} = p_x, p_y$.[3] Reproduced with permission from the American Physical Society.

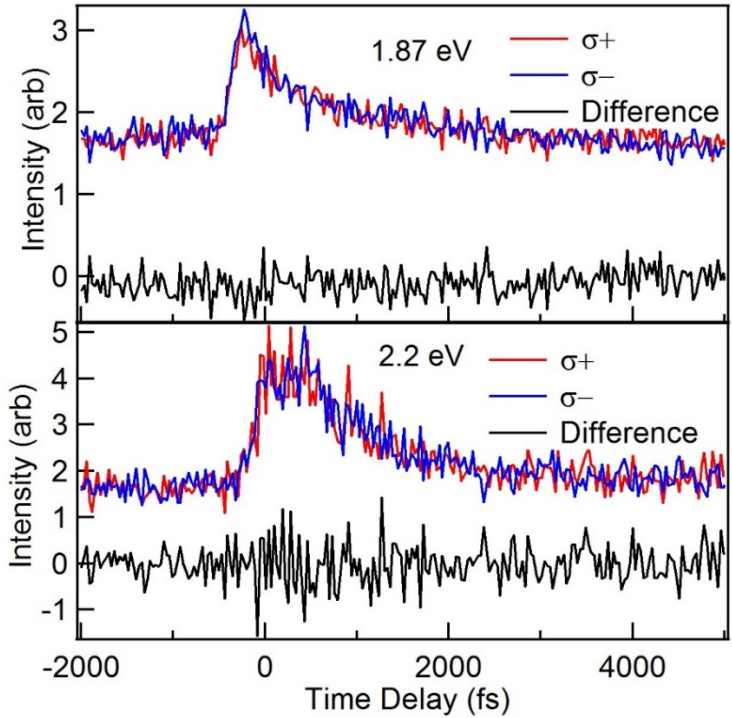

Figure S9. Time resolved CB electron signal intensities collected at K point, with right and left circularized pump. The valley polarization cannot be resolved with either 2.2 eV and 1.87 eV pump photon energies.

Experimental Methods

<u>Sample preparation</u>

MoS$_2$ and WS$_2$ monolayers with mm-cm lateral sizes using a metal mediated exfoliation technique from bulk MoS$_2$ (SPI Supplies) and WS$_2$ (HQ graphene) single crystal, as detailed in ref. [27]. For grounding in photoemission experiments, a 150 nm thick gold film is evaporated onto two sides of the heterostructures. The crystal orientations of the MoS$_2$ and WS$_2$ monolayers is determined using second harmonic generation (SHG), as shown in Fig. S1. The sample is annealed at 250 °C in UHV ($10^{-10}$ Torr) for 4h before photoemission experiments are carried out.

Time and angle-resolved photoemission spectroscopy (TR-ARPES)

A detailed description of the visible pump - EUV probe TR-ARPES setup is published previously.[4] Briefly, The TR-ARPES measurements are carried out in the ultrahigh vacuum chamber ($\sim 10^{-10}$ Torr) at room temperature. The EUV at 22 eV generated with high harmonic generation (KMLabs, XUUS4, modified for 400 nm pump) from the second harmonic of a Ti:sapphire amplifier (Coherent Legend Elite Duo HE+, 10 kHz, 800 nm, 10 W, 35 fs) are used as probe. The visible output form a home-built noncolinear optical parametric amplifier (NOPA, 500-650nm, 35 fs, pumped with the same Ti:sapphire amplifier) is used as a pump. The photoemitted electrons are measured on a hemispherical electron energy analyzer equipped with a 2D delay line detector (SPECS Phoibos-100).